\documentstyle[preprint,aps,psfig]{revtex}
\begin{document}
\tolerance=10000
\draft
\title{Absorption of a Randomly Accelerated Particle:
Gambler's Ruin in a Different Game}
\author{Dominique J. Bicout$^1\,$ and Theodore W. Burkhardt$^{*}$}
\address{$^1\,$ INFM-Operative group Grenoble CRG IN13\\
Institut Laue-Langevin, B.P. 156, F-38042 Grenoble Cedex 9,
France}
\date{\today} 
\maketitle 
\begin{abstract}
We consider a particle which is randomly accelerated by Gaussian
white noise on the line $0<x<1$, with absorbing boundaries at $x=0,1$. 
Denoting the initial
position and velocity of the particle by $x_0$ and $v_0$ and
solving a Fokker-Planck type equation, we derive the exact
probabilities $q_0(x_0,v_0)$, $q_1(x_0,v_0)$ 
of absorption at $x=0,1$, respectively. The results
are in excellent agreement with computer simulations.
\end{abstract}
\vskip 3.0cm  
\noindent Permanent address: Department of Physics, Temple University, 
Philadelphia, PA 19122.
\vfill\eject \narrowtext

A well known topic in random walk theory \cite{f} 
is the problem of the ``gambler's 
ruin.'' Initially the gambler has an amount of money $x_0$ and the bank
the amount $1-x_0$. The gambler flips a coin repeatedly, randomly winning 
or losing the increment $\epsilon$. The game ends when the gambler's funds 
reach $0$ or $1$. The problem is to compute the probability $q_0(x_0)$ 
that the gambler loses everything.
 
The problem is easily solved. 
Since $q_0(x_0)={1\over 2}
[q_0(x_0+\epsilon)+q_0(x_0-\epsilon)]$,
\begin{equation}
{d^2 q_0(x_0)\over dx_0^2}=0\label{a1}
\end{equation}
in the limit $\epsilon\to 0$.
>From equation (\ref{a1}) and the boundary conditions
$q_0(0)=1$, $q_0(1)=0$,
\begin{equation}
q_0(x_0)=1-x_0.\label{a2}
\end{equation}
As the starting capital increases from $0$ to $1$, the probability
of the gambler's ruin decreases from $1$ to $0$. 

Instead of the gambling scenario one could equally well imagine a particle 
making a random walk with infinitesimal steps $\pm\epsilon$
on the $x$ axis, with initial position $0<x_0<1$. In the course of time
the particle eventually arrives at $x=0$ or $x=1$. The quantities
$q_0(x_0)$ in equation (\ref{a2}) and $q_1(x_0)=1-q_0(x_0)$
represent the probabilities that the particle 
first reaches the edge of the interval at $x=0$ and $x=1$, respectively.
Alternatively, we could impose absorbing boundary conditions 
and interprete $q_0(x_0)$ and $q_1(x_0)$ as the probabilities of
absorption at $x=0$ and at $x=1$.

In this Letter we also consider a particle on
the finite interval $0<x<1$, but we assume that the changes in the 
{\em velocity} rather than the position of the particle are random. 
The particle moves according to the Langevin equation
\begin{equation}
{d^2x\over dt^2}=\eta (t),\label{a3}
\end{equation}
where the acceleration $\eta (t)$ has the form of Gaussian white noise, with 
\begin{equation}
\langle\eta(t)\rangle =0,\quad 
\langle\eta(t_1)\eta(t_2)\rangle = 2\delta(t_1-t_2).\label{a4} 
\end{equation}
Imposing absorbing boundary conditions, we derive the probabilities 
$q_0(x_0,v_0)$, $q_1(x_0,v_0)$ of absorption at $x=0$ and 
at $x=1$, respectively, as functions of the initial position and 
velocity. 

The quantity $q_0(x_0,v_0)$ can also be interpreted as the probability of 
a gambler's ruin, but the game is different. The gambler has an 
amount of money $x(t)$ at time $t$ and the bank the amount $1-x(t)$. 
Money is transferred 
from the bank to the gambler at a rate $v=dx/dt$, which may be positive 
or negative. At regular infinitesimal intervals the gambler flips a coin, 
randomly increasing or decreasing the rate $v$ by 
the increment $\Delta$. The game ends 
when $x$ reaches $0$ or $1$. The quantity $q_0(x_0,v_0)$ 
is the probability that a gambler with initial conditions $x_0,v_0$ 
loses everything.

In the case of a random walk on the $x$ axis, the probability density
$P(x,x_0,t)$ at time $t$ of a particle which is initially at $x_0$
obeys the diffusion equation. For a particle which is randomly accelerated
according to equations (\ref{a3}) and (\ref{a4}), the probability
density $P(x,v;x_0,v_0;t)$ in the phase space $(x,v)$ satisfies 
the Fokker-Planck equation \cite{fp}
\begin{equation}
\left(\;{\partial\over\partial t}+v{\partial\over\partial x}-
{\partial^2\over\partial v^2}\right)
P(x,v;x_0,v_0;t)=0\label{a5},
\end{equation} 
corresponding to diffusion of the velocity, with initial condition 
\begin{equation}
P(x,v;x_0,v_0;0)=\delta(x-x_0)\delta(v-v_0).\label{a6}
\end{equation}

In analogy with the discussion leading to differential equation
(\ref{a1}) for $q_0(x_0)$, let us consider a discrete dynamics in
which the velocity $v$ changes by
$\pm\Delta$ with equal probability at time intervals $\tau$. For
this dynamics
\begin{eqnarray}
&&P(x,v;x_0,v_0;t)={1\over 2}[P(x-v\tau,v+\Delta;x_0,v_0;t-\tau)+
P(x-v\tau,v-\Delta;x_0,v_0;t-\tau)],\label{a7}\\
&&q_0(x_0,v_0)={1\over 2}[q_0(x_0+v_0\tau,v_0+\Delta)+
q_0(x_0+v_0\tau,v_0-\Delta)].\label{a8}
\end{eqnarray} Expanding equations (\ref{a7}) and (\ref{a8}) in $\tau$ and 
$\Delta$, dividing by $\tau$, and taking the limit $\tau={1\over 2}
\Delta^2\to 0$ gives us a ``poor man's'' derivation
of the Fokker-Planck equation (\ref{a5}) and the corresponding 
differential equation
\begin{equation}
\left(\;v_0{\partial\over\partial x_0}+{\partial^2\over\partial v_0^2}\right)
q_0(x_0,v_0)=0\label{a9}
\end{equation} 
for the probability of absorption at $x=0$.

To solve equation (\ref{a9}) with the 
absorbing boundary condition
\begin{equation}
q_0(0,v_0)=1,\quad v_0<0,\label{a10}
\end{equation}
and the requirements
\begin{eqnarray}
&&q_0(x_0,v_0)=q_1(1-x_0,-v_0)\label{a11}, \\
&&q_0(x_0,v_0)+q_1(x_0,v_0)=1,\label{a12}
\end{eqnarray}
of reflection symmetry and total probability equal to $1$,
we first make the substitution
\begin{equation}
\psi(x,v)=q_0(x,-v)-{1\over 2}.\label{a13}
\end{equation}
Expressed in terms of $\psi(x,v)$, equations 
(\ref{a9})-(\ref{a12}) take the form
\begin{eqnarray}
&&\left(\;v{\partial\over\partial x}-{\partial^2\over\partial v^2}\right)
\psi(x,v)=0,\label{a14}\\
&&\psi(0,v)={1\over 2},\quad v>0,\label{a15}\\
&&\psi(x,v)=-\psi(1-x,-v).\label{a16}
\end{eqnarray}

Masoliver and Porr\`a \cite{mp} have shown how certain
Fokker-Planck type equations on the finite interval
$0<x<1$ can be solved exactly. They derived an exact result for
the average time $T(x_0,v_0)$ a randomly accelerated particle 
with initial conditions $x_0,v_0$ 
takes to reach a boundary of the interval. 
The probability that 
the particle has not yet reached a boundary after a time $t$ decays
as $e^{-Et}$, as discussed by Burkhardt \cite{twb2}. He 
obtained $E$ numerically with an approach similar to
\cite{mp} and related it to the confinement free energy of a 
semiflexible polymer in a tube. In another application inspired
by \cite{mp}, Burkhardt, Franklin, and Gawronski
\cite{bfg} calculated the equilibrium distribution function
$P(x,v)$ of a randomly accelerated particle on the line $0<x<1$ 
undergoing inelastic collisions at the boundaries \cite{csb}.

The function $\psi(x,v)$ satisfies the same steady-state 
Fokker-Planck equation (\ref{a14}) as the quantity $P(x,v)$ 
considered in \cite{bfg} and has the same Green's function solution
\begin{eqnarray}
\psi(x,v) & = & \frac{v^{1/2}}{3x}\int_0^\infty du\, 
u^{3/2}e^{-(v^3+u^3)/9x}\;I_{-1/3}\left({2v^{3/2}u^{3/2}\over
9x}\right)\psi(0,u) \nonumber\\
 & & -\frac{1}{3^{1/3}\Gamma(\frac{2}{3})}
\int_0^x dy\,\frac{e^{-v^3/9(x-y)}}{(x-y)^{2/3}}\,{\partial
\psi(y,0)\over\partial v},\quad v>0\label{a17}
\end{eqnarray} derived in \cite{bfg}. Equation (\ref{a17}) only
holds for positive $v$. For negative $v$, $\psi(x,v)$ can be obtained
from equation (\ref{a17}) using the antisymmetry (\ref{a16}) under
reflection.

Equation (\ref{a17}) determines $\psi(x,v)$
for all $x>0$ and $v>0$ from $\psi(0,v)$
and $\partial\psi(x,0)/\partial v$. The first of these functions
is given in equation (\ref{a15}). To determine the second, we 
set $v=0$ in equation (\ref{a17}), which yields
\begin{equation}
\psi(x,0)={1\over 3^{1/3}\Gamma({2\over 3})}\left[x^{-2/3}\int_0^\infty du
\thinspace ue^{-u^3/9x}\psi(0,u)-\int_0^x {dy\over (x-y)^{2/3}}\thinspace
{\partial\psi(y,0)\over\partial v}\right].\label{a18} 
\end{equation}
Then, substituting equation (\ref{a18}) in the relation 
$\psi(x,0)+\psi(1-x,0)=0$, 
which follows from (\ref{a16}), and using $\partial\psi(y,0)/\partial v=
\partial\psi(1-y,0)/\partial v$, also a consequence of (\ref{a16}),
we obtain
\begin{equation}
\int_0^1 {dy\over|x-y|^{2/3}}\thinspace {\partial\psi(y,0)\over\partial v}=
\int_0^\infty du\thinspace u\left[{e^{-u^3/9x}\over x^{2/3}}
+{e^{-u^3/9(1-x)}\over (1-x)^{2/3}}\right]\psi(0,u).\label{a19}
\end{equation} 
The solution to integral equation (\ref{a19}), derived, following 
\cite{ps}, in Appendix B of \cite{bfg}, is given by
\begin{equation}
{\partial \psi(x,0)\over\partial v}=\int_0^\infty du
\;u\left[R(x,u)+R(1-x,u)\right]\psi(0,u),\label{a20}
\end{equation}
where
\begin{equation}
R(x,u) = \frac{1}{3^{5/6}\Gamma({1\over 3})\Gamma({5\over 6})}\,\,
\frac{u^{1/2}e^{-u^3/9x}}{x^{7/6}(1-x)^{1/6}}\,\,
 _1F_1\left(\textstyle{-{1\over 6}},{5\over 6},
\frac{u^3(1-x)}{9 x}\right),\label{a21}
\end{equation}
and $_1\thinspace F_1(a;b;z)$ is the confluent 
hypergeometric function \cite{gr,as}.

Equations (\ref{a17}),(\ref{a20}), and (\ref{a21}) 
determine $\psi(x,v)$ for 
all $x$ and $v$ from $\psi(0,v)$ for $v>0$, which is known from the absorbing
boundary condition (\ref{a15}). Substituting equations
(\ref{a15}) and (\ref{a21})
in (\ref{a20}) leads to
\begin{equation}
{\partial\psi(x,0)\over\partial v}={1\over 3^{1/6}\Gamma({1\over 3})}
[x(1-x)]^{-1/6},\label{e1}
\end{equation}
and from (\ref{a15}),(\ref{a17}), and (\ref{e1})
\begin{equation}
\psi(x,v)={1\over 2}-{1\over 2\pi}\int_0^{x}dy\thinspace 
{e^{-v^3/9(y-x)}\over (y-x)^{2/3}}\thinspace [y(1-y)]^{-1/6}.
\label{e2}
\end{equation}
Rewriting equation (\ref{e2}) in terms of $q_0(x_0,v_0)$ 
using (\ref{a11})-(\ref{a13}), we obtain our main result
\begin{equation}
q_0(x_0,v_0)=1-q_0(1-x_0,-v_0)={1\over 2\pi}\int_{x_0}^1 dy\thinspace 
{e^{-v_0^3/9(y-x_0)}\over (y-x_0)^{2/3}}\thinspace [y(1-y)]^{-1/6},
\quad v_0>0,\label{a22}
\end{equation}
analogous to the solution (\ref{a2}) of the traditional 
gambler's ruin problem.

For $x_0=1$ equation (\ref{a22}) reproduces the expected result 
$q_0(1,v_0)=1-q_0(0,-v_0)=0$, $v_0>0$, corresponding to the immediate 
absorption of a particle that is initially at either boundary with velocity
directed outward from the interval $0<x<1$. For $x_0=0$ and $v_0=0$ the 
integral in equation (\ref{a22})
can be evaluated, yielding
\begin{eqnarray}
q_0(0,v_0)&=&1-q_0(1,-v_0)=1-{2^.3^{2/3}\over\Gamma(\textstyle{1\over 6})}
\thinspace 
v_0^{1/2}\, _1F_1(\textstyle{1\over 6};{7\over 6};-{1\over 9}v_0^3),
\quad v_0>0,\label{a23}\\
q_0(x_0,0)&=&1-q_0(1-x,0)=1-{6\Gamma({1\over 3})\over\Gamma({1\over 6})^2}
\; x_0^{1/6}\,
_2F_1(\textstyle{1\over 6},{5\over 6};{7\over 6};x_0).\label{a24}
\end{eqnarray}
Here $_1F_1(a;b;z)$ and $_2\thinspace F_1(a,b;c;z)$ are the 
confluent and ordinary hypergeometric functions \cite{gr,as}.

The probability $q_0(x_0,v_0)$ of absorption at the origin, obtained 
from equation (\ref{a22}) by numerical integration, is shown 
in figure 1. The probability decreases monotonically as $x_0$ increases with 
fixed $v_0$ and as $v_0$ increases at fixed $x_0$, as expected.
The quantity $q_0(x_0,v_0)$ is a nonsingular function of $(x_0,v_0)$ except 
at the two boundary points $(0,0)$ and $(1,0)$. The curves for $x_0=
0.0,0.1,0.3,0.5$ become 
smoother near $v_0=0$ as $x_0$ increases,
and for $x_0=0.5$, $q_0(x_0,v_0)-{1\over 2}$
is an odd function of $v_0$, as implied by equations 
(\ref{a11}) and (\ref{a12}). 

The points in figure 1 show the results of computer simulations,
which clearly are in excellent agreement with the analytical results.
Our simulation routine is based on the exact
solution \cite{gb}
\begin{equation}
P_{\rm free}(x,v;x_0,v_0;t)=\frac{\sqrt{3}}{2\pi t^2}\,
\exp\left\{-\frac{3}{t^3}\,\left[(x-x_0-v_0t)(x-x_0-vt)+
\frac{1}{3}(v-v_0)^2t^2\right]\right\}\label{a25}
\end{equation}
of the Fokker-Planck equation (\ref{a5}) with initial condition (\ref{a6}) 
in the absence of boundaries.
Trajectories with the probability distribution 
$P_{\rm free}(x_{n+1},v_{n+1};x_n,v_n;\Delta_{n+1})$ given by (\ref{a25})
are generated using the algorithm
\begin{eqnarray}
x_{n+1}&=&x_n+v_n\Delta_{n+1}+\left({\Delta_{n+1}^3\over 6}\right)^{1/2}\,
\left(s_{n+1}+\sqrt{3}\,r_{n+1}\right),\label{a26} \\
v_{n+1}&=&v_n+\left(2\Delta_{n+1}\right)^{1/2}\,r_{n+1},\label{a27}
\end{eqnarray}
where $x_n$ and $v_n$ are the position 
and velocity of the particle at time $t_n$, and
$\Delta_{n+1}=t_{n+1}-t_n$. The quantities 
$r_n$ and $s_n$ are independent Gaussian random 
numbers such that 
\begin{equation}
\langle r_n\rangle=\langle s_n\rangle=0,\quad
\langle r_n^2\rangle=\langle s_n^2\rangle=1.\label{a28}
\end{equation}
In the absence of boundaries there is no time-step error in the algorithm, 
i.e., the $\Delta_n$ may be chosen
arbitrarily. Close to boundaries small time steps are needed.

To derive a quantitative criterion for an acceptable time step, 
we begin with the averages 
\begin{equation}
\langle x(t)\rangle=x_0+v_0t,\quad 
\langle[x(t)-\langle x(t)\rangle]^2\rangle
={2\over 3}t^3,\label{a29}
\end{equation}
implied by the distribution function (\ref{a25}).
At time $t$ the particle coordinate $x$ has a Gaussian distribution,
with a maximum at $x=x_0+v_0t$ and the root-mean-square width 
$\left({2\over 3}t^3\right)^{1/2}$. The effect of 
the boundaries on the propagation is
negligible if the Gaussian peak lies almost entirely within the 
interval $0<x<1$. This is certainly the case 
if, say,
\begin{equation}
0<x_0+v_0t\pm 5t^{3/2}<1\label{a30}
\end{equation}
Over the range of velocities encountered in our simulations, any $t$ which
satisfies the simpler, more stringent condition 
\begin{equation}
t<{1\over 10}x_0(1-x_0)\label{a31}
\end{equation}
also satisfies (\ref{a30}).

Keeping inequality (\ref{a31}) in mind, we performed our simulations with the 
time step
\begin{equation}
\Delta_{n+1}=10^{-5}+10^{-1}x_n(1-x_n).\label{a32}
\end{equation}
The time step decreases as the particle approaches the boundary and has
the minimum value $10^{-5}$. It is necessary to have a small nonzero minimum 
value. Otherwise the particle never arrives at the boundaries.
Our results for the absorption probability $q_0(x_0,v_0)$ 
are averages based on $10^5$ trajectories for each set of initial conditions
$x_0,v_0$.

Finally we note that $q_0(x_0,v_0)$
may be derived from another general
Green's function solution 
\footnote{This solution may be derived by
slightly modifying the derivation in Appendix A of \cite{bfg}.
Setting $v=0$ in equation (A3) of \cite{bfg}, solving for $W(s)$, 
and reinserting the
result in (A3) with $v\neq 0$ yields the Laplace transform
of the new solution (\ref{a33}).} of the Fokker-Planck 
equation (\ref{a14}),
\begin{eqnarray}
\psi(x,v) & = & \frac{v^{1/2}}{3x}\int_0^\infty du\, 
u^{3/2}e^{-(v^3+u^3)/9x}\;I_{1/3}\left({2v^{3/2}u^{3/2}\over
9x}\right)\psi(0,u) \nonumber\\
 & & +\frac{v}{3^{2/3}\Gamma(\frac{1}{3})}
\int_0^x dy\,\frac{e^{-v^3/9(x-y)}}{(x-y)^{4/3}}\,
\psi(y,0),\quad v>0,\label{a33}
\end{eqnarray}
different from (\ref{a17}). 
By substituting equation (\ref{a33}) 
in (\ref{a14}), one can check that the Fokker-Planck 
equation is indeed satisfied. On the lines $x=0$
and $v=0$ equation (\ref{a33}) reduces to the 
identities $\psi(0,v)=\psi(0,v)$
and $\psi(x,0)=\psi(x,0)$, respectively.

Equation (\ref{a33}) determines $\psi(x,v)$
for all $x>0$ and $v>0$ from $\psi(0,v),v>0$
and $\psi(x,0)$. The first of these
functions is given in equation (\ref{a15}). To determine the second, we
differentiate equation (\ref{a33}) with respect to $v$ and 
then set $v=0$, which yields
\begin{eqnarray}
&&{\partial\psi(x,0)\over\partial v}={1\over 3^{2/3}\Gamma({1\over 3})}
\times\nonumber\\
&&\left[x^{-4/3}\int_0^\infty du
\thinspace u^2 e^{-u^3/9x}\psi(0,u)-3x^{-1/3}\psi(0,0)
-3\int_0^x {dy\over (x-y)^{1/3}}\thinspace
{\partial\psi(y,0)\over\partial y}\right].\label{a34} 
\end{eqnarray}
For the absorbing boundary condition (\ref{a15}) the first two terms on the
right-hand side of (\ref{a34}) cancel. Substituting equation (\ref{a34}) 
in the relation $\partial\psi(x,0)/\partial v
-\partial\psi(1-x,0)/\partial v=0$, which follows from (\ref{a16}),
using the invariance of $\partial\psi(y,0)/\partial y$ under $y\to 1-y$, 
and integrating with respect to $x$ 
yields
\begin{equation}
\int_0^1 dy\thinspace |x-y|^{2/3} 
{\partial\psi(y,0)\over\partial y}={\rm const}.\label{a35}
\end{equation} 
The function $\psi(x,0)$ given in equations (\ref{a13}) and (\ref{a24})
satisfies equation (\ref{a35}). Substituting this
$\psi(x,0)$ and $\psi(0,v)={1\over 2}$ into equation (\ref{a33}),
integrating, and using (\ref{a11})-(\ref{a13}), we obtain
\begin{eqnarray}
q_0(x_0,v_0)&=&1-q_0(1-x_0,-v_0)\nonumber\\
&=&{2^. 3^{1/3}v\over \Gamma({1\over 6})^2}
\int_{x_0}^1 dy\thinspace {e^{-v_0^3/9(y-x_0)}
\over (y-x_0)^{4/3}}\thinspace (1-y)^{1/6}\,
_2F_1(\textstyle{1\over 6},{5\over 6};{7\over 6};1-y),\quad v_0>0.\label{a36}
\end{eqnarray}
With the help of the identity
\begin{equation}
\int_0^x dy\thinspace{e^{-v^3/9(x-y)}\over (x-y)^{2/3}}\thinspace f(y)=
{v\over 3^{2/3}\Gamma({1\over 3})}
\int_0^x dz\thinspace{e^{-v^3/9(x-z)}\over (x-z)^{4/3}}
\int_0^z dy\thinspace{f(y)\over (z-y)^{2/3}}\label{a37}
\end{equation}
for arbitrary $f(y)$, one can convert
expression (\ref{a36}) for $q_0(x_0,v_0)$ to
the simpler form (\ref{a22}). 

The second of the two Green's function solutions (\ref{a17}),(\ref{a33}) 
looks simpler than the first, since no derivatives of $\psi$ appear on the 
right-hand side, but our main result (\ref{a22}) for
$q_0(x_0,v_0)$ is obtained more easily from (\ref{a17}).

\noindent{\bf Acknowledgements}\\
\indent It is a pleasure to thank Bernard Derrida and George W. Ford
for stimulating discussions. TWB appreciates the hospitality of 
the Ecole Normale Sup\'erieure, Paris
and the Institut Laue-Langevin, Grenoble
and was also supported by Temple University.

\begin{figure}[ht]
\vspace{0.3cm}
\centerline{\psfig{figure=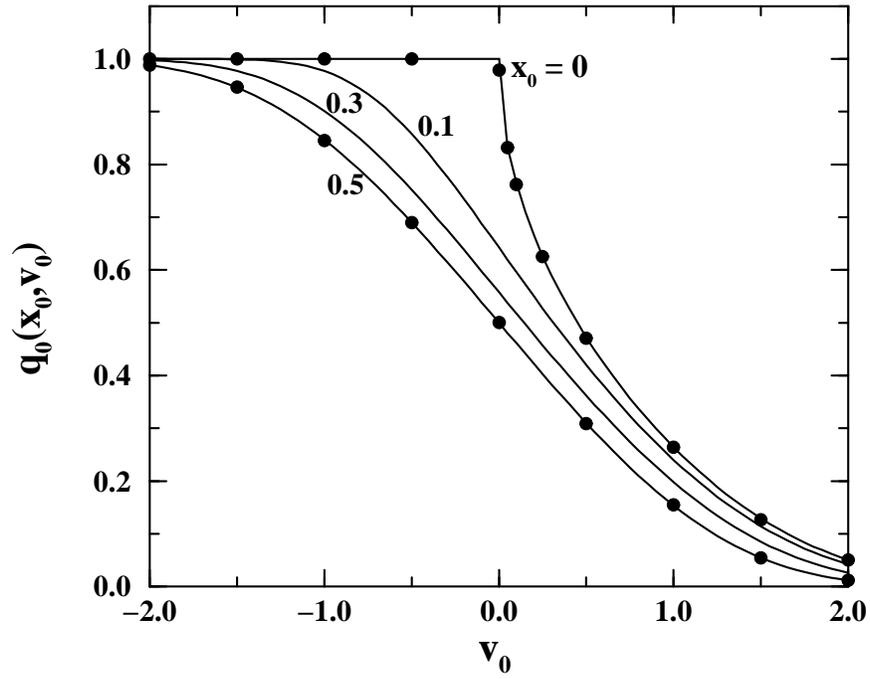,width=4.5in,angle=0}}
\vspace{0.25cm}
\caption{
\noindent The probability $q_0(x_0,v_0)$ of absorption
at the origin. The solid lines show the exact result given in 
equation (\ref{a22}). The points are the results of our computer 
simulations. The data points have a statistical uncertainty
$\pm\delta q_0$ with $|\delta q_0|\stackrel{<}{\sim}0.001$.}
\end{figure}

\begin{references}
\bibitem{f}{Feller W 1968 {\it An Introduction to Probability Theory
and Its Applications, 3d Edition} vol 1 (New York: Wiley 1968)}
\bibitem{fp}{Risken H 1989 {\it The Fokker-Planck Equation: Method 
of Solution and Applications} (Berlin: Springer-Verlag)} 
\bibitem{mp}{Masoliver J and Porr\`a J M
1995 {\it Phys. Rev. Lett.} {\bf 75} 189; 1996 {\it Phys. Rev. E} 
{\bf 53} 2243}
\bibitem{twb2}{Burkhardt T W 1997 {\it J. Phys. A: Math. Gen.} {\bf 30}, 
L167}
\bibitem{bfg}{Burkhardt T W, Franklin J and Gawronski R R 2000
{\it Phys. Rev. E} {\bf 61}, 2376}
\bibitem{csb}{Cornell S J, Swift M R and Bray A J 1998
{\it Phys. Rev. Lett.} {\bf 81} 1142} 
\bibitem{ps}{Porter D and Stirling D S G 1990 {\it Integral Equations} 
(Cambridge: Cambridge University Press)}
\bibitem{gr}{Gradshteyn I S and Ryzhik I M 1980 {\it Tables of Integrals,
Series, and Products} (New York: Academic)}
\bibitem{as}{Abramowitz M and Stegun I A 1965 {\it Handbook of 
Mathematical Functions} (New York: Dover)}
\bibitem{gb}{Gompper G and Burkhardt T W 1989 {\it Phys. Rev. A} 
{\bf 40} 6124} 
\end{references}
\end{document}